\documentclass[twocolumn,prd,nofootinbib]{revtex4}
\usepackage{graphicx}
\usepackage{dcolumn}
\usepackage{amsfonts}
\usepackage{amssymb}
\usepackage{amsmath}

\def\be{\begin{equation}}
\def\ee{\end{equation}}
\def\ba{\begin{eqnarray}}
\def\ea{\end{eqnarray}}

\begin{document}

\title{Minimally coupled scalar fields as imperfect fluids}
\author{Cec\'{\i}lia Gergely, Zolt\'{a}n Keresztes, L\'{a}szl\'{o} \'{A}rp%
\'{a}d Gergely}
%\date{\today }
\affiliation{Institute of Physics, University of Szeged, D\'{o}m t\'{e}r 9, Szeged 6720, Hungary}

\begin{abstract}
We revisit the issue of the fluid description of minimally coupled scalar
fields. While in a cosmological setup the interpretation of a time-evolving
scalar field as a perfect fluid is well-understood, the situation is more
intricate when the scalar field is static, but has a spatial gradient, a
situation motivated by black hole perturbations in scalar-tensor theories.
Then the scalar field is interpreted as either a particular imperfect fluid
of type I or a superposition of a pair of leftgoing (incoming) and
rightgoing (outgoing) null dusts with a perfect fluid. Finally, when the
scalar gradient is null, it is equivalent to an imperfect fluid of type II,
degenerating into null dust when the energy conditions are imposed. We also
propose the suitable action in terms of the fluid pressure components for
each case and discuss the variational principle for a generic class of
minimally coupled scalar fields.
\end{abstract}

\maketitle

\section{Introduction}

Scalar fields recurrently show up in modern gravitational physics either as
generating inflation in the early universe, emerging from dimensional
reduction of higher-order theories, as models of dark matter and dark energy
or as additions to the tensorial sector in modified gravitational models
representing low-energy approximations of the sought for quantum gravity
theory.

At a classical level the most general scalar-tensor theory with at most
second order dynamics both for the scalar and the tensor (hence avoiding
Ostrogradski instabilities) was proposed by Horndeski \cite{Horndeski} and
rediscovered later in the context of generalized galileons \cite{Deffayet}.
In certain higher order theories the degrees of freedom still evolve
according to a second order dynamics, as the analysis of cosmological
perturbations in an effective field theory (EFT) approach has proved \cite%
{GLPVprl,GLPV}. With cosmological symmetries the scalar is purely
time-dependent, hence (provided its gradient never vanishes and it is
timelike) it can be promoted to a time coordinate (unitary gauge).

Odd sector perturbations of spherically symmetric, static black holes in
generic scalar-tensor theories were also discussed in the EFT framework \cite%
{KGT}. Instead of the Arnowitt--Deser--Misner (ADM) decomposition explored
in the cosmological case, here a similar 2+1+1 decomposition \cite{s+1+1a,
s+1+1b} of the 4-metric $\tilde{g}_{ab}$ turned useful: 
\begin{equation}
\tilde{g}_{ab}=-n_{a}n_{b}+m_{a}m_{b}+g_{ab}~,  \label{gtildef1}
\end{equation}%
with $g_{ab}$ a 2-metric on a surface with spherical topology and its
normals satisfying 
\begin{eqnarray}
-n_{a}n^{a} &=&m_{a}m^{a}=1~,  \notag \\
n_{a}m^{a} &=&n^{a}g_{ab}=m^{a}g_{ab}=0~.  \label{norm1}
\end{eqnarray}%
In this case the scalar is static, but has a radial, space-like gradient. If
the latter is nowhere vanishing, the scalar can emulate a radial coordinate
(radial unitary gauge). A nonorthogonal 2+1+1 decomposition was recently
worked out allowing for an unambiguous gauge choice \cite{GKG}, the closest
to the Regge--Wheeler gauge of general relativity, paving the road for the
discussion of the even sector perturbations in an EFT approach of
spherically symmetric, static black holes.

Gravitational dynamics is obtained by varying the action both with respect
to the (inverse) metric tensor and scalar. At first order the respective
equations are generalizations of the Einstein and Klein--Gordon equations.
Second order variations provide the evolutions of perturbations. When other
matter fields are present, their dynamics arises from similar matter field
variations.

The coupling of the scalar to the tensor sector is frame dependent.
Horndeski theories are naturally written up in Jordan frame. In this case
the diffeomorphism-invariant action%
\begin{equation}
S_{G}\left( g^{ab},\phi \right) +S_{\phi }\left( g^{ab},\phi \right)
+S_{M}\left( g^{ab},\psi \right)
\end{equation}%
exhibits minimal coupling of matter fields $\psi $ to the metric, however
the coupling of the scalar field $\phi $ is minimal only in $S_{\phi }$, not
in $S_{G}$. Due to diffeomorphism invariance of the action and the matter
equations of motion the energy-momentum tensor of matter 
\begin{equation}
\tilde{T}_{ab}^{M}=-\tilde{T}_{ab}^{\phi }=-\frac{2}{\sqrt{-g}}\frac{\delta
S_{M}}{\delta g^{ab}}
\end{equation}%
has vanishing covariant 4-divergence \cite{WALD}, nevertheless the
energy-momentum tensor of the scalar field (as defined in terms of the
variation of $S_{\phi }$ with respect to the inverse metric) does not obey
such a conservation law. The scalar however belongs to the gravitational
sector, for the tensorial part of which no proper energy-momentum tensor can
be defined either. Generalised Brans--Dicke type theories have been studied
in this framework \cite{Pimentel,FaraoniCote}, and the imperfect fluid
description of a different, effective energy-momentum tensor worked out. The
relation between the two types of energy-momentum tensors has been discussed
in Ref. \cite{SantiagoSilbergleit}.

By contrast, in the Einstein frame, obtained by a convenient conformal
rescaling of the metric the scalar becomes minimally coupled while the
matter sources cease to be coupled minimally to the metric. In this case the
energy-momentum tensor of matter has no vanishing covariant 4-divergence,
while the energy-momentum tensor of the scalar field has. If no other matter
source is present but the scalar is treated as matter, the Einstein frame is
more natural. When the scalar field is coupled minimally in the Jordan
frame, obviously the two frames and the corresponding energy-momentum tensor
definitions for the scalar coincide.

In the simplest case of general relativity and minimally coupled
Klein--Gordon scalar field with timelike gradient $\tilde{\nabla}_{a}\phi $
(where $\tilde{\nabla}$ is the connection compatible with the metric $\tilde{%
g}_{ab}$), the energy-momentum tensor of the scalar has been shown to mimic
a perfect fluid \cite{Madsen}:%
\begin{equation}
\tilde{T}_{ab}^{PF}=\rho ^{PF}n_{a}n_{b}+p^{PF}\left(
m_{a}m_{b}+g_{ab}\right) ~,  \label{PF}
\end{equation}%
with energy density 
\begin{equation}
\rho ^{PF}=-\frac{1}{2}\tilde{\nabla}_{c}\phi \tilde{\nabla}^{c}\phi
+V\left( \phi \right) ~,  \label{rho}
\end{equation}%
isotropic pressure%
\begin{equation}
p^{PF}=-\frac{1}{2}\tilde{\nabla}_{c}\phi \tilde{\nabla}^{c}\phi -V\left(
\phi \right) ~,  \label{p}
\end{equation}%
and fluid 4-velocity%
\begin{equation}
n_{a}=\frac{\tilde{\nabla}_{a}\phi }{\sqrt{-\tilde{\nabla}_{c}\phi \tilde{%
\nabla}^{c}\phi }}~.  \label{nphi}
\end{equation}

For perfect fluids both $L_{1}=p^{PF}$ and $L_{2}=-\rho ^{PF}$ (regarded as
functions of particle number density and entropy per particle) are valid
Lagrangians \cite{Brown}, differring only by surface terms. However when the
density and pressure become functions of the scalar field, Faraoni has
proven \cite{Faraoni} that the equations of motion of the scalar-tensor
theory are reproduced solely from the Lagrangian (\ref{p}). Related to the
perfect fluid interpretation Ref. \cite{DT} proved that for a finite period
of time a shift-invariant scalar field accurately describes the potential
flow of an isentropic perfect fluid.

The case of a scalar with spatial gradient was deemed nonphysical in Ref. 
\cite{Faraoni} and discussed only briefly (a sign flip being overlooked in
the 3+1 decomposition with respect to the scalar gradient). The correct
expressions for the scalar energy density and isotropic pressure were given
by Ref. \cite{Semiz}, noting that the perfect fluid interpretation does not
hold for a comoving, but rather for a tachyonic observer, when the scalar
gradient is spacelike.

From a timelike observer point of view it is more natural to consider a
2+1+1 fluid decomposition with respect to both the timelike observer and a
preferred spacelike direction,%
\begin{equation}
\tilde{T}_{ab}^{IPF}=\rho n_{a}n_{b}+p_{r}m_{a}m_{b}+p_{t}g_{ab}~,
\label{IPF}
\end{equation}%
similar to the metric decomposition (\ref{gtildef1}). Such an imperfect
fluid could have different radial and tangential pressures $p_{r}$ and $p_{t}
$. (They are dubbed radial and tangential for simplicity - the decomposition
also applies for different scenarios.) In Section \ref{Perf} we explore this
approach, describing all cases in terms of an imperfect fluid, regardless
whether the gradient of the scalar is timelike, spacelike or null. In the
latter case we need to add heat flow too. The energy conditions will also be
discussed here, also we interpret the scalars with radial gradients as a sum
of perfect fluid, incoming and outgoing null dust. The case of null gradient
reduces to a null dust, after the energy conditions are imposed. Finally we
revisit the issue of the proper action for a wide class of minimally coupled
scalar fields in Section \ref{Lag}. We summarize our findings in the
Conclusion.

Throughout the paper we assume $16\pi G=1=c$.

\section{Klein--Gordon scalar field in general relativity as imperfect fluid 
\label{Perf}}

\subsection{Timelike scalar field gradient}

By adding the Klein--Gordon Lagrangian (\ref{p}) to the Einstein--Hilbert
action, (inverse) metric variation yields the energy momentum tensor 
\begin{equation}
\tilde{T}_{ab}^{KG}=\tilde{\nabla}_{a}\phi \tilde{\nabla}_{b}\phi -\tilde{g}%
_{ab}\left[ \frac{1}{2}\tilde{\nabla}_{c}\phi \tilde{\nabla}^{c}\phi
+V\left( \phi \right) \right] ~  \label{EnMom}
\end{equation}%
for the scalar field. For a timelike scalar gradient, as discussed before,
this mimics the perfect fluid (\ref{PF})-(\ref{p}) \cite{Madsen,Faraoni}.

The energy-momentum tensor being diagonal (type I) for both timelike and
spacelike scalar field gradients, the energy conditions translate to $\rho
\geq 0$, $\rho +p_{\alpha }\geq 0$ (weak), $\rho \geq 0$, $\left\vert
p_{\alpha }\right\vert \leq \rho $ (dominant) and $\rho +p_{\alpha }\geq 0$, 
$\rho +\sum_{\alpha }p_{\alpha }\geq 0$ (strong). For timelike scalar
gradient these imply $\rho \geq 0$ (thus $-\tilde{\nabla}_{c}\phi \tilde{%
\nabla}^{c}\phi \geq -2V\left( \phi \right) $) for the weak energy
condition, $V\left( \phi \right) \geq 0$ (then the weak energy condition $%
\rho \geq 0$ is fulfilled automatically) for the dominant energy condition,
finally $-\tilde{\nabla}_{c}\phi \tilde{\nabla}^{c}\phi \geq V\left( \phi
\right) $ for the strong energy condition. All energy conditions hold for $%
0\leq V\left( \phi \right) \leq $ $-\tilde{\nabla}_{c}\phi \tilde{\nabla}%
^{c}\phi $.

\subsection{Spacelike scalar field gradient}

For a spacelike $\tilde{\nabla}_{a}\phi $ the energy-momentum is rather of
the form of an imperfect fluid (\ref{IPF}). The scalar field gradient then
is associated to the spatial vector%
\begin{equation}
m_{a}=\frac{\tilde{\nabla}_{a}\phi }{\sqrt{\tilde{\nabla}_{c}\phi \tilde{%
\nabla}^{c}\phi }}~.  \label{mphi}
\end{equation}%
Then $T_{ab}m^{a}m^{b}$ gives the radial pressure%
\begin{equation}
p_{r}=\frac{1}{2}\tilde{\nabla}_{a}\phi \tilde{\nabla}^{a}\phi -V\left( \phi
\right) ~,  \label{pr}
\end{equation}%
while the tangential pressure and energy density are identified as%
\begin{equation}
p_{t}=-\rho =-\frac{1}{2}\tilde{\nabla}_{a}\phi \tilde{\nabla}^{a}\phi
-V\left( \phi \right) ~.  \label{pt}
\end{equation}

For spacelike scalar gradient the energy conditions are as follows: $\rho
\geq 0$ (thus $\tilde{\nabla}_{c}\phi \tilde{\nabla}^{c}\phi \geq -2V\left(
\phi \right) $) for the weak energy condition, $V\left( \phi \right) \geq 0$
(then the weak energy condition $\rho \geq 0$ is fulfilled automatically)
for the dominant energy condition, finally $V\left( \phi \right) \leq 0$ for
the strong energy condition. All energy conditions hold only for vanishing
potential.

By imposing all energy conditions 
\begin{equation}
\rho =p_{r}=-p_{t}=\frac{1}{2}\tilde{\nabla}_{a}\phi \tilde{\nabla}^{a}\phi
>0~
\end{equation}%
emerges, representing an imperfect fluid with radial pressure equaling its
energy density, and tangential tension of the same magnitude.

Two equivalent interpretations will be discussed below.

\subsubsection{Perfect fluid as seen by a tachyonic observer}

The energy density (8) and isotropic pressure (5) of Ref. \cite{Semiz},
emerging from the perfect fluid interpretation of the scalar by a tachionic
observer 
\begin{equation}
\tilde{T}_{ab}^{tach}=\left( \rho ^{tach}+P^{tach}\right) m_{a}m_{b}+P^{tach}%
\tilde{g}_{ab}~,
\end{equation}%
relate to the radial and tangential pressures of the anisotropic fluid as 
\begin{equation}
\rho ^{tach}=p_{r}-2p_{t}~,\quad P^{tach}=p_{t}~.
\end{equation}%
This interpretation advanced in Ref. \cite{Semiz} is less attractive due to
the nonexistence of tachyonic observers.

\subsubsection{Incoming and outgoing radiation fields superposed on a
perfect fluid}

The imperfect energy-momentum tensor of the scalar field with spatial
gradient,%
\begin{equation}
\tilde{T}_{ab}^{IPF}=-p_{t}n_{a}n_{b}+p_{r}m_{a}m_{b}+p_{t}g_{ab}~,
\end{equation}%
with $p_{r}$ and $p_{t}$ given by Eqs. (\ref{pr})-(\ref{pt}) can be
rewritten as a sum of a perfect fluid (with energy density $-p_{r}$ and
isotropic pressure $p_{t}$)%
\begin{equation}
\tilde{T}_{ab}^{\left( 1\right)
}=-p_{r}n_{a}n_{b}+p_{t}m_{a}m_{b}+p_{t}g_{ab}~,
\end{equation}%
and two null dusts (with the same energy density $p_{r}-p_{t}$):%
\begin{equation}
\tilde{T}_{ab}^{\left( 2\right) }=\left( p_{r}-p_{t}\right)
k_{a}k_{b}~,\quad \tilde{T}_{ab}^{\left( 3\right) }=\left(
p_{r}-p_{t}\right) l_{a}l_{b}~.
\end{equation}%
The null dusts propagate in the null directions 
\begin{equation}
k_{a}=\frac{n_{a}+m_{a}}{\sqrt{2}}~,\quad l_{a}=\frac{n_{a}-m_{a}}{\sqrt{2}}%
~,
\end{equation}%
which span a pseudoorthonormal basis obeying 
\begin{eqnarray}
k_{a}k^{a} &=&l_{a}l^{a}=0~,\quad k_{a}l^{a}=-1~,  \notag \\
k^{a}g_{ab} &=&l^{a}g_{ab}=0~.
\end{eqnarray}%
The null dusts represent leftgoing (incoming for spherical symmetry) and
rightgoing (outgoing) radiation fields. Such null dusts add up to an
anisotropic fluid with no tangential pressures \cite{Letelier}, a scenario
explored for describing a static superposition of incoming and outgoing
radiations under spherical symmetry \cite{2ND}. A similar dynamical
construction under spatial homogeneity yielded a Kantowski--Sachs type
homogeneous universe filled by a two component radiation, evolving from an
initial singularity to a final one \cite{2NDhom}. Switching to ghost
radiation streams (negative energy densities) the corresponding solutions
represented wormholes \cite{Hayward}, naked singularities or open universes 
\cite{worm}. These all emerged in a dilatonic approach as solutions for the
massless scalar field minimally coupled to the spherically reduced
Einstein--Hilbert gravity. In our case the addition of a\ third, perfect
fluid is necessary to account for the tangential pressure.

By imposing all energy conditions, the perfect fluid will have negative
energy density (ghost fluid) and negative isotropic pressure (tension), both
equaling $-\rho $, while the energy density of both null dusts simplifies to 
$2\rho $.

\subsection{Null scalar field gradient\label{null}}

Finally we discuss the case when $\tilde{\nabla}_{a}\phi $ is null. Then we
rather decompose the metric into the pseudoorthonormal basis:%
\begin{equation}
\tilde{g}_{ab}=-2k_{(a}l_{b)}+g_{ab}~,
\end{equation}%
consistent with the metric decomposition (\ref{gtildef1}).

Next we associate $k_{a}$ with the scalar field as 
\begin{equation}
k_{a}=\frac{\tilde{\nabla}_{a}\phi }{\sqrt{2}}~.  \label{kphi}
\end{equation}%
Then the energy-momentum tensor (\ref{EnMom}) simplifies:%
\begin{equation}
\tilde{T}_{ab}^{KG}=2k_{a}k_{b}-\tilde{g}_{ab}V\left( \phi \right) ~,
\label{Tnull}
\end{equation}%
while in the $\left( n^{a},m^{a}\right) $ basis it reads%
\begin{eqnarray}
\tilde{T}_{ab}^{KG} &=&n_{a}n_{b}+2n_{(a}m_{b)}+m_{a}m_{b}+  \notag \\
&&\left( n_{a}n_{b}-m_{a}m_{b}-g_{ab}\right) V\left( \phi \right) ~,
\end{eqnarray}%
or in matrix form%
\begin{equation}
\tilde{T}_{ab}^{KG}=\left( 
\begin{array}{ccc}
1+V\left( \phi \right) & 1 &  \\ 
1 & 1-V\left( \phi \right) &  \\ 
&  & -g_{ab}V\left( \phi \right)%
\end{array}%
\right) ~.  \label{Tnulldust}
\end{equation}%
This energy-momentum tensor is of Type II, according to the classification
of Ref. \cite{HawkingEllis}, as it should be due to the double eigenvector $%
k^{a}$ of the energy-momentum tensor (\ref{Tnull}).

For the weak energy condition the tangential pressures should be positive,
hence $V\left( \phi \right) \leq 0$, but on the other hand the energy
density should be $\geq 1$, thus $V\left( \phi \right) \geq 0$. Hence the
weak energy condition holds only for $V\left( \phi \right) =0$. A similar
conclusion stems from imposing either the dominant or the strong energy
condition. With $V\left( \phi \right) =0$ the scalar field becomes a
massless radiation field (null dust). As shown in Ref. \cite{Ivanov}, and
transparent from Eq. (\ref{Tnulldust}) with $V\left( \phi \right) =0$ the
null dust can also be perceived as an imperfect fluid with energy density,
radial pressure and heat flow, all of them equal.

\section{Lagrangian description as imperfect fluid of a minimally coupled
generic scalar field\label{Lag}}

The proof in Ref. \cite{Faraoni} that the correct Lagrangian in the case of
timelike scalar field gradient is the isotropic pressure (\ref{p}), combined
with the anisotropic fluid description applying for the case of a spatial
gradient implies that the Lagrangian in the latter case is the tangential
pressure (\ref{pt}). The same expression vanishes in the case of a scalar
with null gradient obeying the energy conditions. This case will be
discussed in the more generic framework below.

The result that the isotropic pressure of a perfect fluid mimicking the
scalar field with timelike gradient qualifies as Lagrangian applies to a
much wider class of minimally coupled scalar-tensor theories:%
\begin{equation}
L=L\left( X,\phi \right) ~,  \label{lag}
\end{equation}%
with%
\begin{equation}
X=\frac{1}{2}\tilde{\nabla}_{a}\phi \tilde{\nabla}^{a}\phi ~.
\end{equation}%
(For the Klein--Gordon field $L=X-V$.) Variation with respect to $\phi $
gives the dynamics of the scalar field:%
\begin{equation}
\nabla _{a}\left( L_{X}\tilde{\nabla}^{a}\phi \right) =L_{\phi }\left( \phi
\right) ~.  \label{scalarEL}
\end{equation}%
Variation with respect to the (inverse) metric results in%
\begin{eqnarray}
\delta S_{\phi } &=&\!\int dx^{4}\!\left[ L\left( X,\phi \right) \delta 
\sqrt{-\tilde{g}}\!+\!\frac{1}{2}\sqrt{-\tilde{g}}L_{X}\tilde{\nabla}%
_{a}\phi \tilde{\nabla}_{b}\phi \delta \tilde{g}^{ab}\right]  \notag \\
&=&\!\!\int dx^{4}\frac{\sqrt{-\tilde{g}}}{2}\!\left[ \!-\!L\left( X,\phi
\right) \tilde{g}_{ab}\!+\!L_{X}\tilde{\nabla}_{a}\phi \tilde{\nabla}%
_{b}\phi \right] \!\delta \tilde{g}^{ab}~
\end{eqnarray}%
(the subscript $X$ on $L$ denoting partial derivative with respect to $X$).
The energy-momentum tensor arises as%
\begin{equation}
\tilde{T}_{ab}^{\phi }=-\frac{2}{\sqrt{-\tilde{g}}}\frac{\delta S}{\delta 
\tilde{g}^{ab}}=-L_{X}\tilde{\nabla}_{a}\phi \tilde{\nabla}_{b}\phi +\tilde{g%
}_{ab}L\left( X,\phi \right) ~.
\end{equation}

\subsection{Timelike scalar field gradient}

For a timelike scalar field gradient $\tilde{\nabla}_{a}\phi $ a 4-velocity $%
n^{a}$ can be associated with the scalar field through Eq. (\ref{nphi}), and
the metric decomposed in the manner (\ref{gtildef1}). This yields the
energy-momentum tensor%
\begin{equation}
\tilde{T}_{ab}^{\phi }=\left( 2XL_{X}-L\right) n_{a}n_{b}+\left(
m_{a}m_{b}+g_{ab}\right) L\left( X,\phi \right) ~,
\end{equation}%
in a form of a perfect fluid with energy density%
\begin{equation}
\rho =2XL_{X}-L\mathcal{~},
\end{equation}%
and isotropic pressure 
\begin{equation}
p=L\mathcal{~}.
\end{equation}%
Thus the Lagrangian density is but the pressure of the fluid.

\subsection{Spacelike scalar field gradient}

For a space-like scalar field gradient $\tilde{\nabla}_{a}\phi $ associated
to the vector $m^{a}$ through Eq. (\ref{mphi}) the energy momentum tensor
rather becomes the imperfect fluid%
\begin{equation}
\tilde{T}_{ab}^{\phi }=\left( L-2XL_{X}\right) m_{a}m_{b}+\left(
-n_{a}n_{b}+g_{ab}\right) L\left( X,\phi \right) ~,
\end{equation}%
with energy density%
\begin{equation}
\rho =-L~,
\end{equation}%
and pressure components%
\begin{equation}
p_{r}=L-2XL_{X}~,
\end{equation}%
\begin{equation}
p_{t}=-\rho =L~.
\end{equation}%
Thus in this case the Lagrangian density is $p_{t}=-\rho $, extending the
result established for the Klein--Gordon field to a generic minimally
coupled scalar.

\subsection{Null scalar field gradient}

For a null scalar field gradient $\tilde{\nabla}_{a}\phi $ associated to the
vector $k^{a}$ through Eq. (\ref{kphi}) $X=k^{a}k_{a}=0$ holds, hence 
\begin{equation}
\tilde{T}_{ab}^{\phi }=-2L_{X}\left( \phi \right) k_{a}k_{b}+\tilde{g}%
_{ab}L\left( \phi \right) ~.
\end{equation}%
A similar analysis to the one presented in Subsection \ref{null} shows that
this is of type II and the energy conditions are satisfied for $L\left( \phi
\right) =0$, rendering the scalar to a null dust\footnote{%
The strategy to follow here is to insert $X=0$ and $L\left( \phi \right) =0$
only in the equations derived from the variational principle, rather than
into the Lagrangian (\ref{lag}). This is how a nonvanishing $L_{X}\left(
\phi \right) $ enters the equations, however $L_{\phi }\left( \phi \right)
=0 $ holds.}.

The diffeomorphism invariance of the scalar energy-momentum tensor (\ref{lag}%
) implies $\tilde{\nabla}^{a}\tilde{T}_{ab}^{\phi }=0$, which (due to $%
\tilde{\nabla}^{a}L_{X}\left( \phi \right) \propto k^{a}$ and $\tilde{\nabla}%
_{b}L\left( \phi \right) =\sqrt{2}L_{\phi }\left( \phi \right) k_{b}$) leads
to a geodesic equation%
\begin{equation}
k^{a}\tilde{\nabla}_{a}k_{b}=\left( \frac{L_{\phi }\left( \phi \right) }{%
\sqrt{2}L_{X}\left( \phi \right) }-\tilde{\nabla}_{a}k^{a}\right) k_{b}~.
\end{equation}%
On the other hand the scalar dynamical equation (\ref{scalarEL}) implies%
\begin{equation}
\tilde{\nabla}_{a}k^{a}=\frac{L_{\phi }\left( \phi \right) }{\sqrt{2}L_{X}}~,
\end{equation}%
hence the geodesic is affinely parametrized:%
\begin{equation}
k^{a}\tilde{\nabla}_{a}k_{b}=0~.
\end{equation}%
Thus the null gradient of the minimally coupled generic scalar field (\ref%
{lag}) obeys an affinely parametrized geodesic equation, similarly as found
for the Klein--Gordon field in Ref. \cite{FaraoniCoteNull}.

\section{Concluding remarks}

We have revisited the fluid description of a minimally coupled scalar field
to gravity. For scalar fields with timelike gradient (in situations
motivated by cosmological considerations) the fluid is perfect, both for the
Klein--Gordon field and for a quite generic scalar with Lagrangian (\ref{lag}%
), with the equation of state provided by the scalar field itself. The
isotropic pressure serves as Lagrangian. However the scalar field has a
spatial gradient when discussing spherically symmetric, static black hole
solutions and their stability in scalar-tensor gravity theories. It has been
known in the literature that sometimes the scalar behaves as a radiation
field.

Hence the cases of scalar fields with spatial or null gradient require
special attention. In both cases the fluid corresponding to the scalar field
is imperfect, but simple enough due to the minimal coupling. In the spatial
case the energy-momentum tensor is diagonal, of type I. The tangential
pressure serves as Lagrangian and (being its opposite) it also determines
the energy density. The radial pressure is different. We have shown that
such an energy-momentum tensor can be equally interpreted as a superposition
of a perfect fluid and a pair of leftgoing (incoming) and rightgoing
(outgoing) radiation streams represented by null dusts.

In the null case the energy-momentum tensor is of type II, representing an
imperfect fluid with different energy density, radial and tangential
pressures, also heat flow.

We discussed the restrictions imposed by the energy conditions in all three
cases. In particular in the spacelike and null cases the energy conditions
switch off the potential, hence the mass of the scalar. In the null case the
scalar field degenerates into a null dust.

It is well-known that massless particles forming a null dust follow null
geodesics. In general these geodesics are not affinely parametrised. The
freedom to rescale null vectors, while the energy density of null dust is
rescaled such to preserve the form of the null dust energy-momentum tensor
allows to achieve the divergenceless of the null vector, hence the affine
parametrisation \cite{BicakKuchar}. By imposing the energy conditions, the
minimally coupled scalar field becomes a null dust, already affinely
parametrized.

Remarkably even without energy conditions fulfilled, a generic minimally
coupled scalar field with null gradient still evolves along affinely
parametrized null geodesics.

\section{Acknowledgements}

This work was supported by the Hungarian National Research Development and
Innovation Office (NKFIH) in the form of the Grant No. 123996 and has been
carried out in the framework of COST actions CA15117 (CANTATA), CA16104
(GWverse) and CA18108 (QG-MM), supported by COST (European Cooperation in
Science and Technology). C.G. was supported by the UNKP-19-3 New National
Excellence Program of the Ministry for Innovation and Technology. Z.K. was
supported by the J\'{a}nos Bolyai Research Scholarship of the Hungarian
Academy of Sciences and by the UNKP-19-4 New National Excellence Program of
the Ministry for Innovation and Technology.

\end{document}